\documentclass[lettersize,journal]{IEEEtran}
\usepackage{amsmath,amsfonts}
\usepackage{algorithm}
\usepackage{algpseudocode}
\usepackage{array}
\usepackage[caption=false,font=normalsize,labelfont=sf,textfont=sf]{subfig}
\usepackage{textcomp}
\usepackage{stfloats}
\usepackage{url}
\usepackage{verbatim}
\usepackage{graphicx}
\usepackage{cite}
\usepackage{array}
\usepackage[inkscapelatex=false]{svg}
\begin{document}

\title{IDEAS: Information-Driven EV Admission in Charging Station Considering  User Impatience to Improve QoS and Station Utilization}
\author{
Animesh Chattopadhyay$^1$,
~\IEEEmembership{Member,~IEEE}
Subrat Kar$^2$,
~\IEEEmembership{Senior Member,~IEEE}
\thanks{The authors are attached to Indian Institute of Technology, Delhi (IIT Delhi), India (email:$^1$animesh.chat@ee.iitd.ac.in; $^2$subrat@ee.iitd.ac.in)}
\thanks{This work has been submitted to the IEEE for possible publication. Copyright may be transferred without notice, after which this version may no longer be accessible.}
}

\maketitle

\begin{abstract}

Our work delves into user behaviour at Electric Vehicle(EV) charging stations during peak times, particularly focusing on how impatience drives \textit{balking} (not joining queues) and \textit{reneging} (leaving queues prematurely).
We introduce an Agent-based simulation framework that incorporates user optimism levels (pessimistic, standard, and optimistic) in the queue dynamics.
Unlike previous work, this framework highlights the crucial role of human behaviour in shaping station efficiency for peak demand.
The simulation reveals a key issue: balking often occurs due to a lack of queue insights, creating user dilemmas.
To address this, we propose real-time sharing of wait time metrics with arriving EV users at the station.
This ensures better Quality of Service (QoS) with user-informed queue joining and demonstrates significant reductions in reneging (up to 94\%) improving the charging operation.
Further analysis shows that charging speed decreases significantly beyond 80\%, but most users prioritize full charges due to range anxiety, leading to a longer queue.
To address this, we propose a two-mode, two-port charger design with power-sharing options.
This allows users to fast-charge to 80\% and automatically switch to slow charging, enabling fast charging on the second port.
Thus, increasing fast charger availability and throughput by up to 5\%.
As the mobility sector transitions towards intelligent traffic, our modelling framework, which integrates human decision-making within automated planning, provides valuable insights for optimizing charging station efficiency and improving the user experience.
This approach is particularly relevant during the introduction phase of new stations, when historical data might be limited.


\end{abstract}

\begin{IEEEkeywords}

Electric Vehicle, Fast Charging, Queuing Theory, Agent-Based Model, Charging Management, QoS, Intelligent Transportation

\end{IEEEkeywords}


\section{Introduction}


\IEEEPARstart{F}{or} a carbon-neutral future, the transition from Internal Combustion Engine (ICE) vehicles to Electric Vehicles (EVs) is a key milestone.
However, EVs need periodic charging and the charging infrastructure must be ubiquitous to reduce range anxiety. 
Good legislation, good policy, and standards help, but the growth in EV charging infrastructure still does not meet the charging demand.
Therefore, the limited charging infrastructure must be used optimally by managing the charging demand of EV users -- using algorithms which efficiently implement policies in pricing, admission control, and service priorities.

Before deploying such algorithms, we need simulation frameworks for accurately modelling Electric Vehicle(EV) user traffic at the charging station(CS).
Typically in conventional simulations, an EV arrives at the CS and joins a queue (single input and single output), then leaves after being partly/fully charged by the charger. 
The average waiting time, queue length, and station service throughput are then evaluated. 
EV user preferences and usage decisions must be modelled faithfully factoring in the human element of indecisiveness - which is most often ignored in previous work~\cite{antoun2021data}.
In a real-world scenario, such indecisiveness causes instability in charging queue operation, and users \textit{balk} - they decide not to join the queue and go elsewhere, sometimes missing an imminent chance or they are impatient and \textit{renege} - they leave after spending some time in the queue, thus wasting a spot.
Modelling such user behaviour is an important differentiating feature unique to our proposed simulation framework.

\begin{figure}[t!]
\includegraphics[width=1\columnwidth]{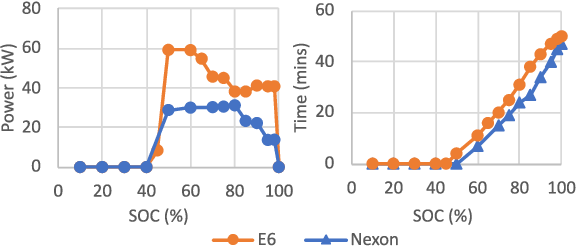}
\caption{Charging Profile (for two popular EV vehicles, BYD's E6 and Tata's Nexon, showing how the charging speed changes with SoC; the speed of charging decreases significantly after 80\%. It takes roughly equal time to charge from 0\% to 80\% and to charge the remaining 20\%\cite{TataNexo73:online}}
\label{fig:chargeProf}
\end{figure}

The time it takes to fully charge an EV is another aspect we have considered.
Fast-charging times have improved from hours to minutes, but due to the inherent charging profile of Li-ion batteries\cite{brighente2022evscout2}, fast charging only happens until 80\% State of Charge (SoC). Beyond 80\%, there is a significant drop in charging speed as evident from cautious
~\ref{fig:chargeProf}.

EV users tend to charge as much as they can to avoid range anxiety.
At peak demand, this practice of full-charging blocks the service for other EVs needing fast charging.
Therefore, admission control strategies are employed to filter out EVs with certain charging demands.
Various dynamic pricing methods and aggressive policies are deployed to discourage EV owners from blocking chargers for a long time -- often leading to a diminished charging experience by the users. 

We have developed a framework to model the behaviour of EV users waiting to be charged, accounting for these human decision-making methods, specifically balking and reneging.
We model a limited waiting space (a finite queue) with options for users to leave without full charging or even without charging at all. 
To assess the loss of revenue-earning charging traffic, the balking customers are also considered. 

Further, we estimate the loss of traffic under two different situations (a) when the charger status information is not available to the incoming EV user and (b) when the incoming EV user is informed about the likely charger status and availability. 
Our results show that to be effectively utilized, situation (b) is better -- the CS should share basic information with potential users. 

We propose that every charger should have two physical ports, each with a fast charge and slow charge mode. 
A port should switch from fast charge to slow charge mode when the SoC reaches 80\%.
We have incorporated this two-mode charger design into our framework.
We show this leads to a significant performance improvement in the throughput of charged users, of up to 5\%.
\begin{figure*}[t]
\includegraphics[width=\textwidth]{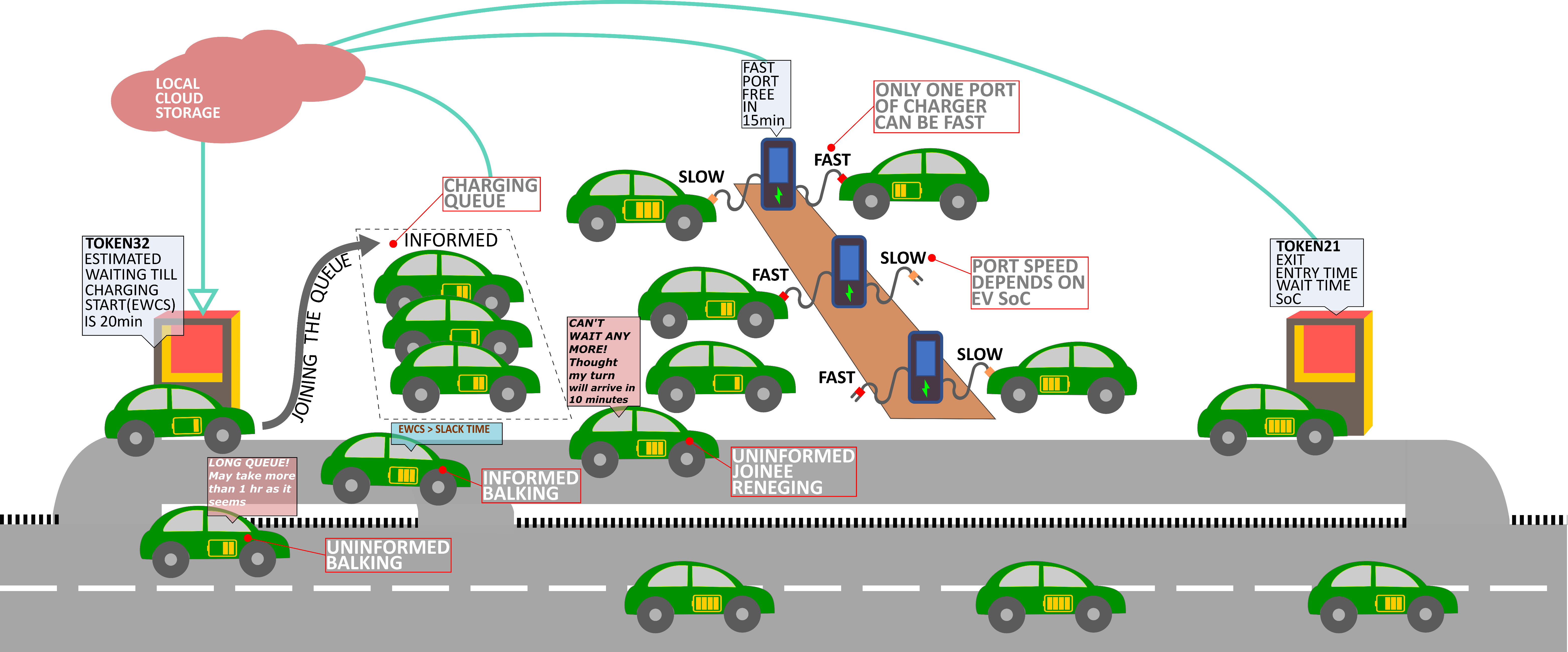}
\caption{The EV charging scenario, illustrating balking and reneging}
\label{fig:illustration}
\end{figure*}

The paper is organized as follows: a survey of existing work and identification of research gap is done in Section \ref{Related Works}.
In Section \ref{formulation}, we show how an M/M/1/k queuing theory model can be extended to make provision for impatient EV users.
Section \ref{simulation} describes the tools and the modelling methods used to build the framework for simulation. 
The results of the simulation are presented in Section \ref{results}, with salient conclusions stated in Section \ref{conclusion}.
%
%
\begin{table}[!t]
\caption{Specifications for some EVs available in India\label{tab:batteryCapacity}}
\centering
\begin{tabular}{|c|c|l|l|}
\hline
EV Name             & Battery   & Mileage                   & Charging Time (SoC)\\
\hline
\hline
TATA Nexon          & 40.5~kWh  & 465~Km                    & 56~min (10\%-80\%) 50~kW DC\\
\hline
BYD E6              & 71.7~kWh  & 520~Km& 1.5~hrs 60~kW DC\\
\hline
MG ZS EV            & 50.3~kWh  & 461~Km                    & 60~min till 80\% 50~kW DC\\
\hline
\end{tabular}
\end{table}

\section{Related Works}\label{Related Works}

There are several existing frameworks \cite{schwenkMultiDayStochasticScheduling2022,zhangSimulatingChargingProcesses2022,tanQueueingNetworkModels2014c,xiaoOptimizationModelElectric2020a} which unlike ours do not focus on the local traffic in the vicinity of the charger, though it is \textit{this} component of traffic which has the most significant impact on the charging demand.
They analyze different charging modes and focus on scheduling EV users to different chargers. 
Some authors \cite{esmailiradExtendedQueueingModel2021,qinChargingSchedulingMinimal2011,wangElectricalVehicleCharging2018,zenginis2016analysis} have simulated the EV charging queue to evaluate the efficiency of the charging strategy; they optimize the profit of the EV CS and fail to incorporate the impatience of the waiting EV users.
Though they consider the waiting-time-in-queue data to distribute EV user traffic, they neither share this with the EV users nor consider user impatience in their results \cite{wang2022queue,schoenbergReducingWaitingTimes2023,zhangDiscreteeventHybridTraffic2019}.

In \cite{moghaddamSmartChargingStrategy2018}, to discourage high demand during peak hours of household electricity consumption, the dynamic electricity cost information is shared with the arriving EV. 
However, their model neglects both impatience and real traffic behaviour, relying instead on heuristics to optimize these objectives.

Zhang et al\cite{zhangDeepLearningBasedProbabilisticForecasting2021b} have tried to forecast the EV charging traffic arrival rate and consider the impatient behaviour of EV drivers -- but they made no provision for the CS to share wait information with the EV users.
The balking is based on the observation of queue length and (generated) random probabilities.
This may make the queue unstable, resulting in significant reneging traffic -- which is not the case in our proposed strategy. 
Lai et al \cite{laiPricingElectricVehicle2023}, \textit{have} considered the impatience of the incoming EV traffic.
However, the loss of traffic was not given priority - a pricing scheme was used instead to discourage the EVs in an aggressive approach to admission control.
The EV user satisfaction was considered in  \cite{ucerModelingAnalysisFast2019,zhangOptimalChargingScheduling2019} where the queue waiting time was estimated and considered as the direct indication of customer satisfaction.
In \cite{zhangOptimalChargingScheduling2019}, the authors considered reneging as the service dropping rate to assess the performance of the CS without considering balking.

In  \cite{antoun2021data}, the average waiting time and probability of reneging are analysed. 
The dataset is not publicly available, and the expression of reneging loss, which is derived from \cite{bocquet2005queueing} is different from our queueing model which is more generalized for handling impatient EV users.
While it does model impatient customers, the results are largely heuristic and are based on charging and battery technology which is dated, and do not suggest any solution to improve the traffic loss or to ensure better QoS or service efficiency.
The charger utilisation is analysed, but the utilisation of parking space goes unaddressed, though it is a more contemporary figure of merit in urban CS infrastructure.

We address these gaps in our simulation framework model, which factors in the impatience of the EV user during charging at peak time. 
Our renege process allows users to leave without receiving service at \textit{any} point of time instead of allowing them to leave only when they reach the head of the queue\cite{antoun2021data,bocquet2005queueing}. 
Unlike others, who admit EVs to a waiting queue based on some admission control (restricting their entry based on demand), we leave it to the EV user to choose to wait(or not wait) in the queue.
We use the estimated waiting time, directly collecting it from the charger and the queue, and share it with the user, to be used for an informed decision to join the queue. 
Our approach optimizes Quality of Service (QoS) without either resorting to aggressive admission policies or discouraging usage through steep pricing.
In addition, we try to increase the availability of the fast charger using our innovative two-mode two-port charging scheme. 
By separating fast charging sessions from slow charging sessions, we free up the fast chargers and increase their availability.

\begin{figure}[t!]
\includegraphics[width=1\columnwidth]{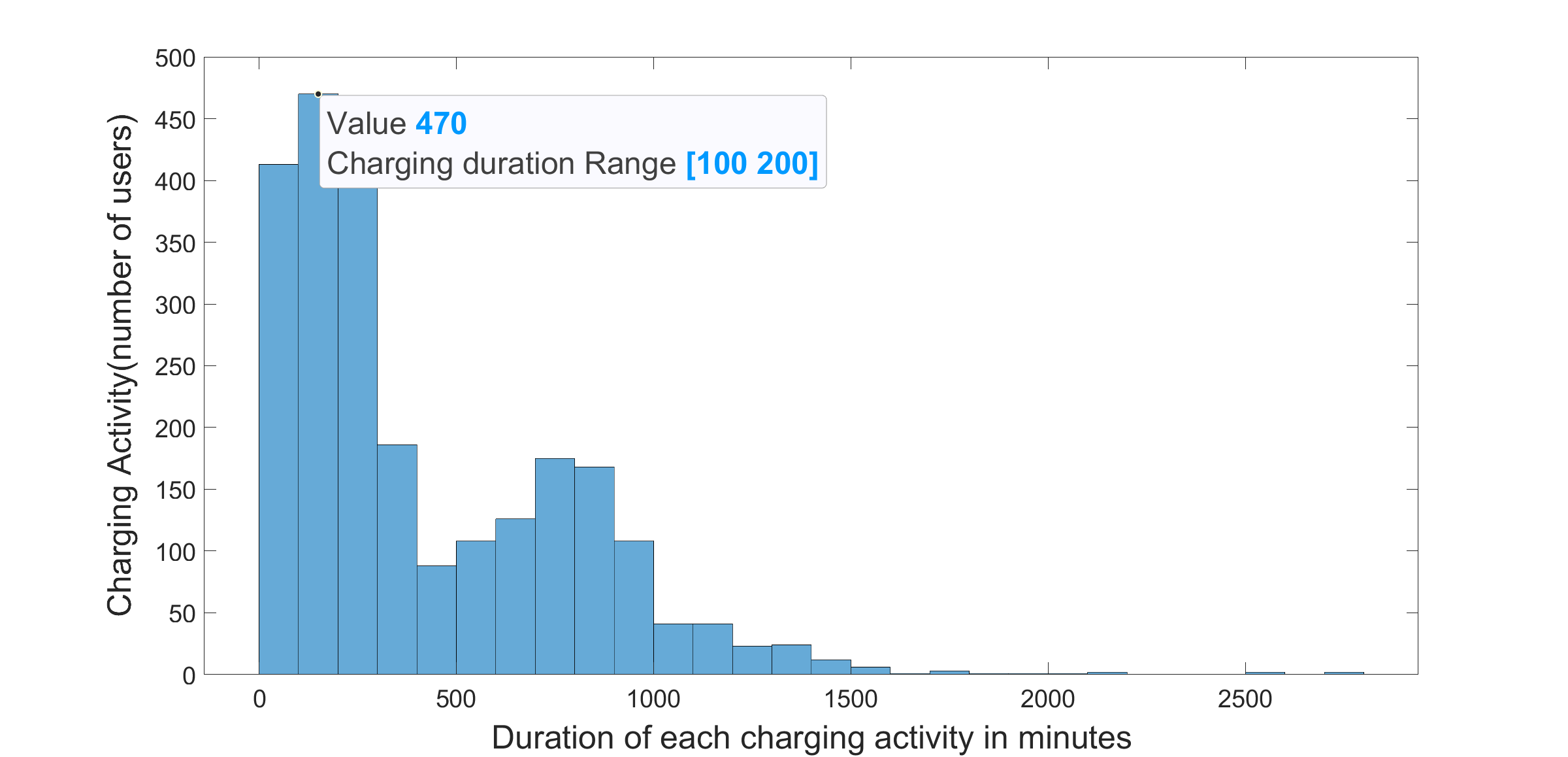}
\caption{Most used charging duration lies between 100 and 200 minutes\cite{lee_acndata_2019}}
\label{fig:chargeDuration}
\end{figure}
\begin{figure}[t!]
\includegraphics[width=1\columnwidth]{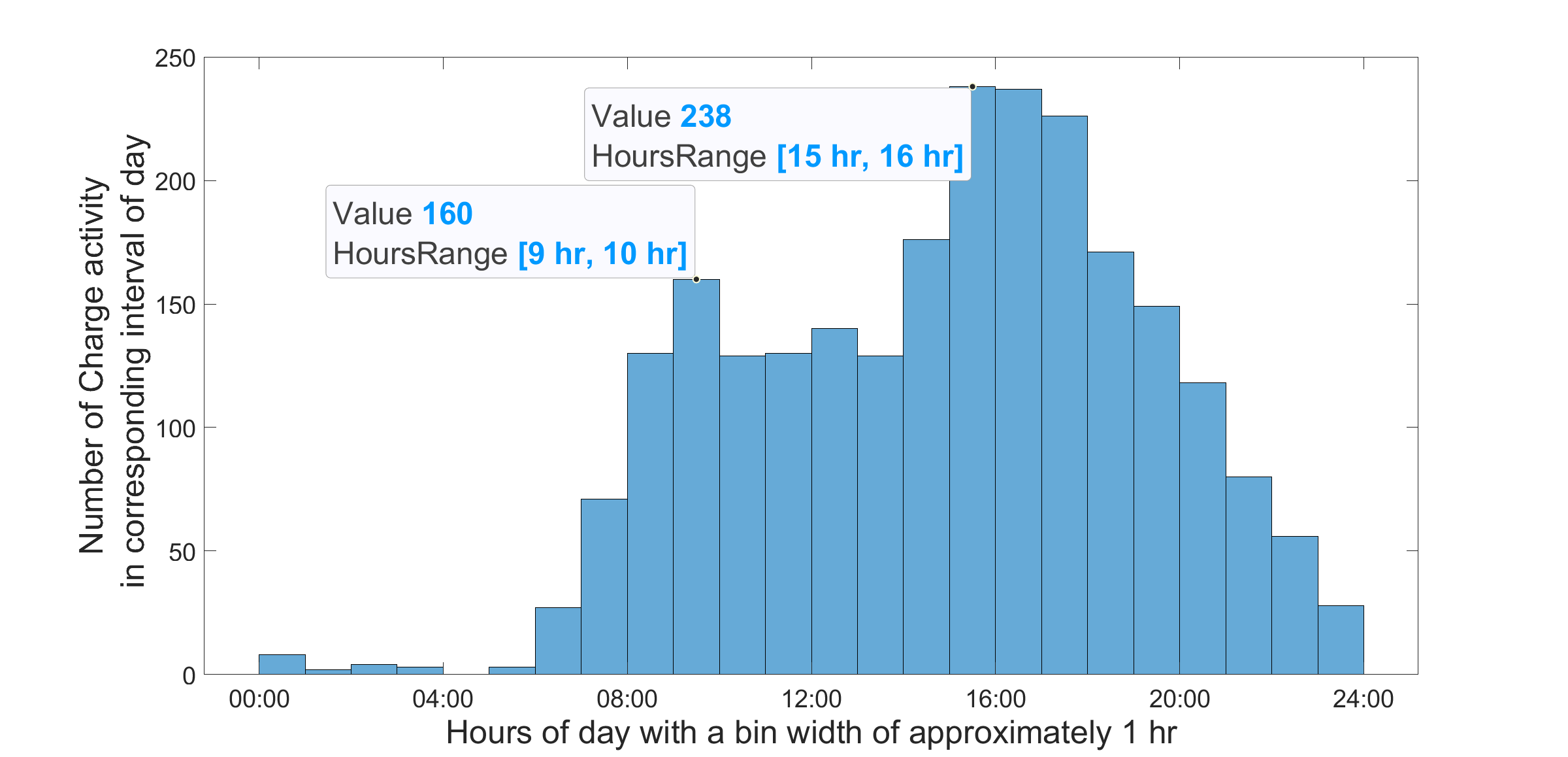}
\caption{Charging traffic distribution over the Day recorded for a period of 6 months-- active hours are between 0700-2300 hours and the demand is higher between 0900-1000 hours and 1400-1900 hours \cite{lee_acndata_2019}}
\label{fig:chargeTraffic}
\end{figure}


\section{Proposed Charging Station Model: Theoretical Overview}\label{formulation}

\begin{figure}[t!]
\includegraphics[width=0.9\columnwidth]{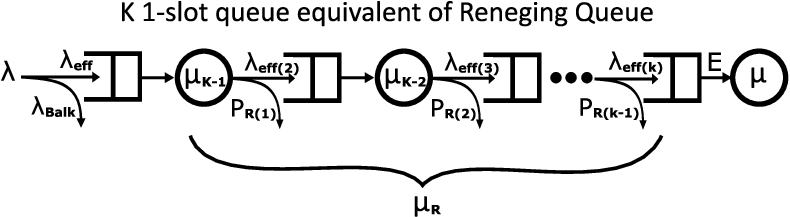}
\caption{Analytical modelling of the Wait Queue with reneging allowed for users}
\label{fig:renegeQueueSplit}
\end{figure}

\subsection{The Queueing Model}
\begin{figure}[t!]
\includegraphics[width=0.9\columnwidth]{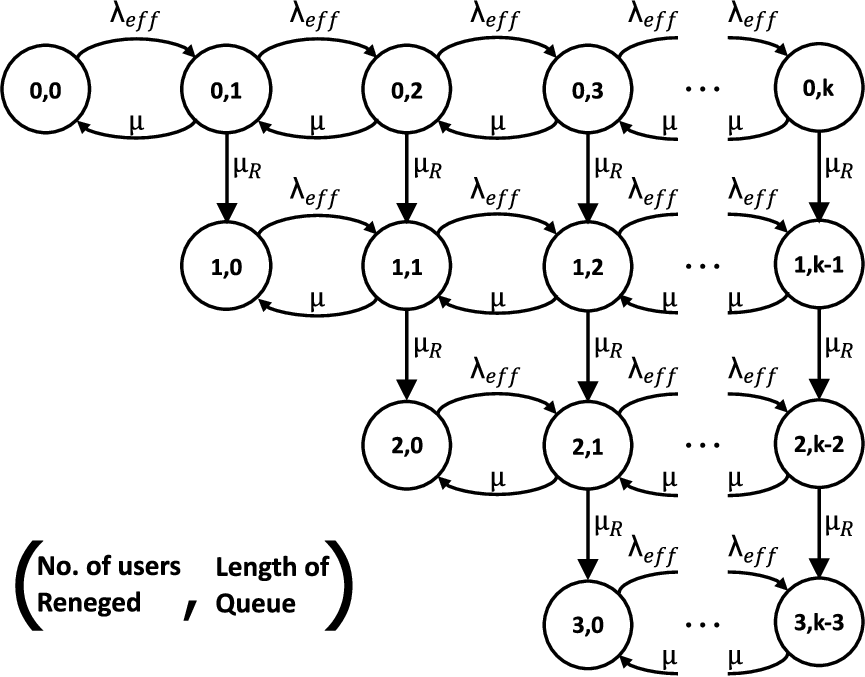}
\caption{Markov chain model of a typical charging station with impatient EV users reneging.} 
\label{fig:Markov}
\end{figure}
We model a single charger station as an M/M/1/k queueing system, where EV arrival follows a Poisson distribution with an arrival rate $\lambda$, as popularly considered in the literature \cite{ucer2019modeling,antoun2021data}
\begin{equation}
\label{eqn:poisson}
    P(\Delta t)=\lambda e^{-\lambda \Delta t}
\end{equation}
Every charger is a server with a service rate $\mu$, exponentially distributed service times, and a waiting queue of size $k$. 

The traffic intensity \(\rho\) is the ratio of arrival rate \(\lambda\) to service rate \(\mu\).
For the stability of the queue, $\rho$ must satisfy the condition $\rho < 1$.
Charging an EV takes a significant amount of time and is a stochastic process, with the EV arriving with a random amount of charge remaining in the battery. At a peak demand time $\lambda >> \mu$ and theoretically, the queue tends to overflow.
In real-world scenarios, typically, the EV users enter the CS, wait in a finite `\'small' ($k$-slot) queue for their turn, and connect to the charger once it is free. 
An EV user with a charging requirement may not join the queue (\textit{Balk}) either if the queue is full (\textit{forced} balking) or if they perceive the queue to be very long and anticipate a waiting time longer than their liking and leave (\textit{voluntary} balking).
Though voluntary balking is very subjective, this can be linked to the intended service time (the amount of charge needed).
The latter is ignored in the literature \cite{antoun2021data} to avoid the complexity of modelling.

If $P_B$ be the probability of balking, the effective rate of arrival into the queue would be 
\begin{equation}
    \lambda_{eff} = \lambda \cdot (1-P_B)
\end{equation}
Here, $\lambda_{eff} < \mu$ .
\textit{Forced} Balking occurs with a probability  $P^F_B$ (Same as a blocking probability $P_b$ \cite{antoun2021data}) when the queue reaches full capacity, thereby preventing new entries. 
This is also equal to the probability $P_k$ that the M/M/1/k system is at full capacity and a $(k+1)$th EV arrives when $k$ EVs are already waiting in the system (and $1$ EV is being charged) \cite{kleinrock1967time}:
\begin{equation}
\label{eq:block}
P^F_B = P_b = P_K=\frac{(1-\rho) \rho^k}{1-\rho^{k+1}} \mathrm{,where~} \rho = \frac{\lambda}{\mu} 
\end{equation}

\textit{Voluntary} Balking occurs with the probability $P^V_B$ when  the user refuses to join because the queue is too long (but not full), and is given by 
$P^V_B =  e^{-(1-w)\sigma}, \sigma \geq 0, 1 \leq w < k$
where $w$ is the number of EVs in the queue 
and $\sigma$ is the parameter that determines how sensitive the Balking probability is to the length of the queue \cite{zhangDeepLearningBasedProbabilisticForecasting2021b}. 
A different derivation of voluntary balking can be found in \cite{sztrik2012basic}.

The probability of Informed Voluntary Balking ($P^{IV}_B$) is defined as a fraction $z$ of the time it will take to charge the \textit{i}th EV from initial SoC when arrived to final SoC, given as
\begin{equation}
\label{eqn:chargeTime}
    T(i,SoC,SoC^f)
\end{equation}
The probability $P_B(\nu)$ that potential 'new' user-to-be  $\nu$ chooses not to join the queue with k waiting users upon arrival based on the available status information, is given as
%
\begin{equation}
\begin{array}{cc}
 \label{eq:InformedVoluntaryBalking}
    P_B(\nu) = &\\   P [ (z \cdot T(\nu,SoC,100\%)) & \\
                 <  (W_q \cdot N(t) 
                     + T(k+1,SoC,80\%) 
               ] &\\
              \end{array}
\end{equation}              
where $z$ = 0.6 is the Impatience Factor, defined as a fraction of the required charging time that the EV user waits for without getting impatient. 
And since there are k users waiting, $T(k+1,SoC,80\%)$ is the {remaining~charge~time} of the EV user $k+1$ who is connected to the charger at time \textit{t}.
The term $z$  shows that the EV user's patience depends on the State of Charge (SoC),  
the expected total waiting time ($W_q$ is the average waiting time per user in the queue and  
$N(t)$ is the instantaneous number of EVs present in the queue at time $t$).

EV users may renege (i.e. leave the queue without charging) when they lose patience after waiting for a while.

Let $\tau$ be the time after which an EV user exits the queue when they are yet to get their chance to charge. 
There are several attempts to study the effect of reneging users \cite{antoun2021data,bocquet2005queueing,sztrik2012basic}.
The threshold time is mostly subjective and varies with EV users' charging goal, patience threshold, and the EV battery SoC.
However, in \cite{antoun2021data} they have adopted the derivation of \cite{bocquet2005queueing} which considers $\tau$ to be a deterministic constant. 
In a more realistic and relatable derivation of reneging \cite{sztrik2012basic}, they considered the reneging rate as $r_k$ in an M/M/c/K queue where
$r_k=(k-c) \theta, r_k=e^{\frac{k \alpha}{c \mu}}, k=c, \ldots K$
\begin{equation}
\mu_k= \begin{cases}k \mu, & k=1, \cdots, c \\ c \mu+r_k, & k=c, \cdots, K\end{cases}
\end{equation}
From the definition, the users are leaving the queue with rate $r_k$ which depends on the queue length and only when all the chargers are occupied. 
Here $\theta$ is the exponentially distributed impatience time of a customer.
The queue architecture, in addition to FIFO propagation, should allow the exit of the EV users from their current position in the queue.
In \cite{antoun2021data} the queue architecture is derived from, \cite{bocquet2005queueing} which considers the exit of the users from the front of the queue near the server(in our model, this is the charger) and also considers a partial service and exit from the server.
To model a more general queue for reneging, we consider the k-slot queue as  \textit{k} 1-slot queues with a propagation server between each, as shown in Figure~\ref{fig:renegeQueueSplit}.
The reneging rate $r_k = \mu_R$.
It is this propagation server which handles the exit of the reneging user when the user either decides to stay in the queue or leaves at the rate $\mu_{K-i}$.
The total population of reneging EV users gives rise to traffic that is served with rate $\mu_R$ towards the exit.
The queue with reneging users can be depicted as a Markov chain, shown in Figure~\ref{fig:Markov} with the states $(a,b)$ where $a$ is the total~number~of~reneged~users and $b$ is the number~of~EVs~waiting~in~the~queue.

The probability of reneging (\(P_r\)) represents the likelihood that a customer, having joined the queue, leaves before receiving service. 
It depends upon the system's reneging policy and the customer's waiting time threshold.

The average number of EVs in the system ($N_{sys}$) considering reneging is the average number of EVs within the system including those in the queue, those being served, and those who have reneged, is given by 
\begin{equation}
\label{renegeLength}
    N_{sys} = \frac{\rho}{1 - \rho} + P_r
\end{equation}

The waiting time $W_q$ for an EV in the queue without receiving service (\(W_q\)) \textit{with reneging}  is 
\begin{equation}
\label{renegeWait}
    W_q = \frac{\rho}{\mu(1 - \rho)} + \frac{P_r}{\lambda_{eff}}
\end{equation}

We start with an M/M/1 queue for simplicity.
In general, however, the charging process is not Markovian and is better modelled using an M/G/1/k queue. 
This makes the theoretical modelling and observation of the system more complex.
There is no closed-form expression for balking and reneging for such a system\cite{antoun2021data}.
Moreover, the $\lambda$ and $\mu$ need to be measured from observation in real-world traffic.
The EV charging data publicly available is mostly outdated and does not reflect the anticipated future demand rise and peak charging times.
Hence, we move on to a simulation method for further analysis and evaluation of our proposed idea.


\section{Proposed Simulation Model}\label{simulation}

In our charging station(CS) model, as given in Figure~\ref{fig:illustration}, the behaviour of EV users mimics real-world scenarios.
Users are free to join (or not join and balk) the charging queue and leave if they lose patience (renege), which is typical of EV user behaviour.
To simulate such a realistic behaviour, we model the EV user as an Agent that can make independent decisions to balk and renege (its user's behavioural characteristics). 

The queue is an integral part of the CS operation, requiring the impatient EV user to renege from their present position in the queue. Hence, we model a modified FIFO queue to enable such an operation in the simulation.

\begin{figure}[!t]
\includegraphics[width=1\linewidth]{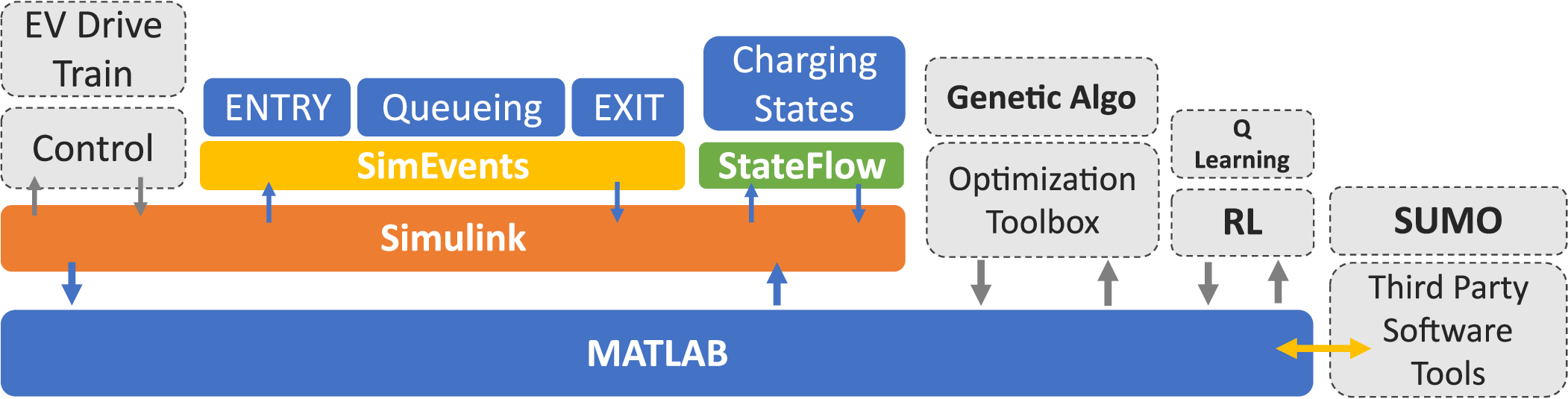}
\caption{Software layers of the Simulation Framework}
\label{fig:layers}
\end{figure}

Our Simulation Framework uses MATLAB and Simulink tools (Figure~\ref{fig:layers}), incorporating multiple programming paradigms from the SimEvents toolbox~\cite{SimEventsUserGuide}. 
Integration with SimEvents enables effective handling of discrete event components, and swift prototyping, as is extensively employed in the traffic management domain \cite{zhangDiscreteeventHybridTraffic2019}. 
The framework allows tailored programming paradigms catering to specific modelling requirements, utilizing Entity Flow, Textual Programming, Graphical Programming, and StateFlow \cite{liExtensibleDiscreteEventSimulation2016a}. 

It operates on event-driven interactions and block behaviours rather than time-based activities, processing entities akin to structured data buses. 
These entities trigger block reactions upon entry and exit events, aided by gates and switches for flow control. 
Notably, SimEvents' crucial functionality lies in queue analysis, providing statistical insights into each block. 

\begin{figure*}[!t]
\includegraphics[width=1\textwidth]{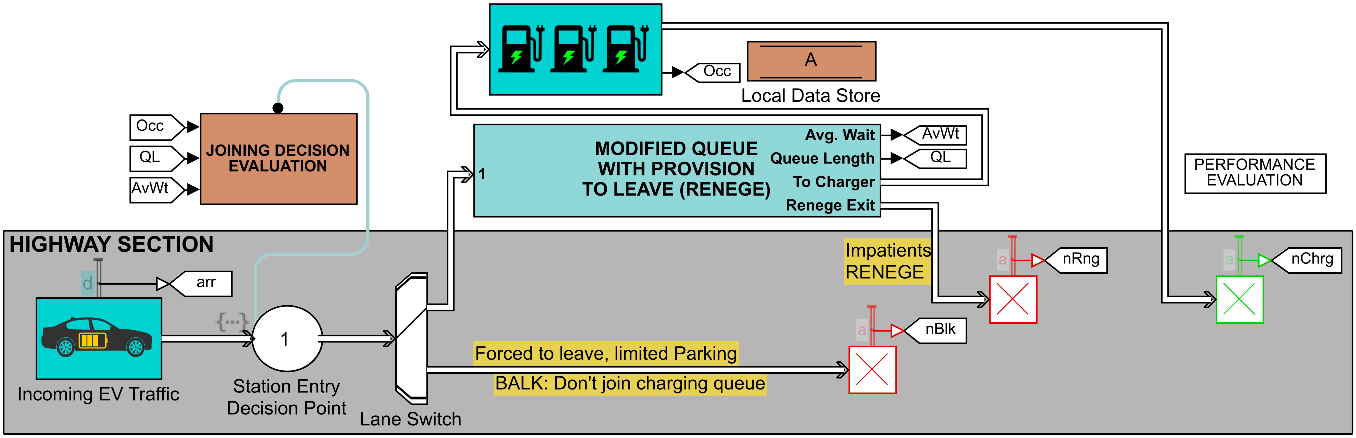}
\caption{Architecture of the Simulation Framework}
\label{fig:simEventsModel}
\end{figure*}

The model of the scenario being simulated is shown in Figure~\ref{fig:simEventsModel}.

\subsection{Behavioral model of the EV User}

The EV agent represents the EV and its user as a single entity with the user's choice and behavioural characteristics.
It borrows from the technical specifications of the popular Indian EV car, the Tata Nexon EV, with its Battery Capacity and Charging Rate as given in Table~\ref{tab:batteryCapacity}.
Additionally, behavioural attributes like patience threshold and charging choice govern how the EV responds in different environments.
In our simulation, the EV agent is exposed to three charging scenarios as discussed in Section~\ref{results}

Within the simulation framework, upon `generation', arriving EVs are initialized with a random SoC drawn from the range 5\%-60\% (assumed uniformly distributed). 
The duration of fast charging depends on the charger's capability.
The expected time to charge from initial SoC till 80\% for the \textit{i}th EV is $T(i,SoC,80\%)$ from Equation~(\ref{eqn:chargeTime})

The \textit{Patience Threshold} is a behavioural attribute of the EV user and represents the maximum waiting time the user can tolerate (in minutes).
This attribute, which enables the EV user to decide whether to join the queue or to leave after an extended wait, is given as 
$\mathrm{Patience~threshold~of~the~ith~impatient~EV}$
\begin{equation}
\label{eqn:impatience}
    T_{Impatience}(i) = z \cdot  T(i,SoC,80\%)
\end{equation} 
where $z=0.6$ from Eqn~(\ref{eq:InformedVoluntaryBalking}).
The lower the initial SoC, the longer the time taken to reach 80\% charge, and more the patience required of the EV owner to remain in the queue.

EV agents respond to various scenarios, acting on stimuli for different decisions and behaviours based on user types.
Three user types, the Optimist, the Standard user and the Pessimist user, were considered in situations where actual data was not available. 
The EV users accounted for scenarios involving forced balking, reneging and some voluntary balking events.

To the Pessimist user $i$ arriving at time $t$, the worst-case waiting time estimate  is important, given by  
\begin{equation}
\label{eqn:OWTS1}
    T^{Waiting}_{Pessimist}(i) 
        = T(i,5\%,80\%) \cdot N_{sys}(t)
\end{equation}
and the estimated Waiting Time for Standard users is 
\begin{equation}
\label{eqn:OWTS2}
    T^{Waiting}_{Standard}(i)
        = T(i,SoC,80\%) \cdot N_{queue}(t)
\end{equation}
and the estimated Waiting Time for Optimistic users is 
\begin{equation}
\label{eqn:OWTS3}
    T^{Waiting}_{Optimist}(i)
        = T(i,60\%,80\%) \cdot N_{queue}(t)
\end{equation}
Where $N_{queue}(t)$ and $ N_{sys}(t)$ are the number of EVs in the queue and system respectively, at time \textit{t}. These estimates are the response times of different user types to stimuli, allowing for scenario-based decisions and reactions based on their inherent traits and charging environment.

The Entity Generator Block(Figure~\ref{fig:simEventsModel}) models the arrival of EV users as a Poisson process.
Entities, consisting of ID, priority, and user-defined data parameters, are used in our modelling to represent EVs.
In our simulation framework, an entity embodies an EV, containing attributes such as the battery's SoC and the user's patience threshold, prompting server actions like charging and reneging events.
The attributes defining the user patience and the EV's battery status are used in the next  Decision Logic Block to decide whether to join or not.

\subsection{Proposed Informed Joining Implementation}

This block simulates a conditional decision logic, according to Eqn~(\ref{eqn:patience_buffer}), that changes according to the cases being simulated.
It simulates the decision taken by the EV user on encountering the CS.
When an arriving user encounters a full queue, blocking occurs, and the EV attributes (e.g. SoC) are ignored.
For modelling an observation (of the queue) based balking decision, the comparison logic uses Eqns~(\ref{eqn:OWTS1} and (\ref{eqn:OWTS2}).
For our proposed informed balking case, this block gathers information about the average wait time from the queue and the time left in charging from the Charger.

To allow the EVs entering the queue to leave without charging, we created a customized queue.

\subsection{The `Impatient' Queue}

The queue, with length $L_Q$ employs a First-In-First-Out (FIFO) discipline to advance EVs. It offers two outputs: one towards the charger (server) and the other to allow EVs to exit when the time already spent waiting exceeds the patience threshold : 
\begin{equation}
\label{eqn:patience_buffer}
\left [ T - T_{time~of~entry}(i) \right ] > T_{Impatience}(i)
\end{equation}

This customized queue, allowing impatient entities to exit, is developed using SimEvents Custom Discrete Event Blocks as in Algorithm~\ref{alg:alg1}. 
\begin{algorithm}[!ht]
\caption{Impatience Monitoring Queue Algorithm}\label{alg:alg1}
\begin{algorithmic}
\State \textbf{Initialize:} Queue Size $L_q$, Number of In Ports I and Number of Out Ports O
\State \textbf{Initialize:} In Ports and Out Ports; 
\State \textbf{ENTRY EVENT: } Entity $E_i$
\State Read attribute SoC, Patience Buffer, $T^{Impatience}(i)$
\State $SoC_{\text{final}}=  SoC_{\text{arrive}} + C_{\text{fast}} \cdot T_f +  {C}_{\text{slow}} \cdot T_s $ \Comment{\%State of charge update process}
\State \textbf{Attach Timer:} $ T^m(i) \gets T^{Impatience}(i) $
\State \textbf{TIMER EVENT:}  $T^m(i)$ Expired
\State \hspace{2em} Forward $EV_i$ to Renege Port
\State \textbf{EXIT EVENT:RENEGE} 
\State \hspace{2em}\textbf{Increment:} Renege Counter
\For{$E_i={1,...,K }$}
\State \textbf{Iterate: Advance Position in Queue} 
\EndFor
\State \hspace{2em}Iterate
\State \textbf{EXIT EVENT:CHARGER} 
\State \hspace{2em}\textbf{Increment:} Service Counter
\For{$E_i={1,...,K }$}
\State \textbf{Iterate: Advance Position in Queue} 
\EndFor

\end{algorithmic}
\end{algorithm}
The queue outputs the average wait time of the block. 
This block also outputs the length of the queue, the average waiting time, and the number of EV users who reneged.

Decision-making processes are influenced by impatience and travel objectives, while flexible queue management permits timely exits. 

Next, to implement our two-state (Fast Charge and Slow Charge) proposed charger architecture, we use StateFlow charts.

\subsection{Proposed EV Fast Charger Implementation}

The proposed two-state charger is efficiently modelled using the StateFlow chart tool to switch between Fast and Slow mode states of charging.
The fast mode charges till 80\% SoC and then the charger switches to Slow mode. 
Thus, it gives the EV user a choice -- to skip the slow charging part(Figure\ref{fig:chargeProf}). 
A cautious user, we expect, would continue to charge till 100\% due to range anxiety (Figure~\ref{fig:chargeDuration}).
The Charger monitors the charging state of the connected EV and sends the information in real-time to the Decision Logic block of the CS to inform the arriving EVs.
We assume that an EV is never unplugged prematurely (before reaching a SoC of 80\% in Fast Charging mode. 
This charger block utilizes features from the State Flow toolbox of MATLAB into SimEvents, and its functionality is elaborated in Algorithm~\ref{alg:alg2}.
\begin{algorithm}[t]
\caption{Two-Mode Two-Port DC Charger Algorithm}\label{alg:alg2}
\begin{algorithmic}
\State \textbf{Declare:} PORT $\gets$ Port1, Port2; ChargeRates $\gets C_{\text{fast}},{C}_{\text{slow}}$

\State \textbf{Define PORT STATE: FAST}
\While {$SoC_{\text{final}} < 80\%$}
\State $SoC_{\text{final}} = SoC_{\text{final}} + C_{\text{fast}} \cdot T_f$
\EndWhile
\State \textbf{Define PORT STATE: SLOW}
\While {$SoC_{\text{final}} \leq 100\%$}
\State $SoC_{\text{final}} = SoC_{\text{final}} + C_{\text{slow}} \cdot T_f$
\EndWhile
\State \textbf{ENTRY EVENT: } Entity $E_i$
\State \textbf{Input}: {$SoC_{\text{arrive}} \gets SoC$\%, Demand in kWh}
\If{(Port1.State == Idle) AND [(Port2.State == Idle) OR (Port1.State == Slow)]}
\State \textbf{PORT} $\gets$ \textbf{Port1}
\State \textbf{PORT.STATE} $\gets$ \textbf{Fast}
\ElsIf{(Port2.State == Idle) AND [(Port1.State == Idle) OR (Port1.State == Slow)]}
\State \textbf{PORT} $\gets$ \textbf{Port2}
\State \textbf{PORT.STATE} $\gets$ \textbf{Fast}
\EndIf
\If{$SoC_{\text{final}} \geq 80\%$}
\State \textbf{PORT.STATE} $\gets$ \textbf{Slow} \Comment{\%Switching Port State}
\EndIf
\State \textbf{EXIT EVENT:} {PORT.STATE} $\gets$ \textbf{Idle} \Comment{\%Free the Port}
\end{algorithmic}
\end{algorithm}


\section{Performance Evaluation}\label{results}

We utilize our custom framework, to model four distinct cases.
Each case is a modification(Case 1 is modified to form Case 2, and so on) and represents a departure from the conventional charging scenario.
In the last one, we model and implement our proposed solution.
We simulate each of these cases to assess the performance of our proposed charging solution.
The four cases are described (each with an acronym by which it is subsequently referred to):
\begin{enumerate}

    \item (\textbf{BlockingFC}) 
    This case describes a scenario where no voluntary balking occurs but there is provision for reneging.
    Arriving EV users are forced to balk when blocked from entering because there is no waiting space (the waiting queue is at full capacity). 

    \item (\textbf{ObservationFC}) 
    In this case, EV users arrive, observe the size of the waiting queue, and then decide whether or not to join the queue, subject to their individual preferences and patience.
    This incorporates voluntary balking based on the waiting time assumption of the arriving EV user after observing the EV population present in the station.
    The EV users renege normally as before upon loss of patience waiting in the queue.
    
    \item (\textbf{InformedFC}) 
    Here, the CS shares the charging status information with the arriving EV users as an estimated wait time.
    The arriving EV users then make an informed decision about whether to join the waiting queue. 
    
    \item (\textbf{Informed2PortCharge}) 
    We provision a two-mode two-port improved charger.
    This allows users to fast-charge to 80\% SoC and automatically the charging port switches to slow mode charging, enabling fast charging on the other port.    
    This allows simultaneous fast charging of another EV while the previous one continues charging until unplugged.
    
\end{enumerate}

The simulation parameters are stated next.

\subsection{Simulation settings}

The arrival of EVs at the station is Poisson distributed as in Equation~(\ref{eqn:poisson}) with $\lambda = 0.6$ for rush hours and $\lambda = 0.1$ for low demand hours.
The initial SoC of arriving EV users is assumed uniformly distributed between a minimum SoC m(5\%) and a maximum SoC M (60\%).
For the {\tt ObservationFC} case. EV users are classified depending on whether their assumptions of wait time is optimistic(low wait), moderate(moderate wait), and pessimistic(long wait).
The patience buffer (or slack time) is $T^{Impatience}(i)$ as in Eqn~(\ref{eqn:patience_buffer}).

We have considered a charging station on the highway with 50kW chargers capable of charging an EV with a 40.5kWh battery from 10\% to 80\% in 56 minutes for reference, as in Table~\ref{tab:batteryCapacity} as per typical state-of-the-art technical specifications.
All the simulation results were obtained after 1000 simulation rounds.
From Figure~\ref{fig:chargeTraffic}, we find two high-demand peaks at 0900hrs and 1500hrs. The active charging hours are from 0700hrs till 2359hrs. We take these 1019 most active minutes into account.

\subsection{Evaluation Metrics}\label{Metrics}
In addition to the conventional metrics of Throughput and Average Waiting Time, We define some new metrics to evaluate the station's performance, considering impatient users.

\textbf{Service Throughput:} The absolute number of EVs served by charging to the desired SoC per unit time.

\textbf{Average Waiting Time:} The Average Time an EV Spends waiting in the Queue

\textbf{Balking \%:} The percentage of EV users who either refuse to join the Queue due to their preference or they leave since the queue is full.

\textbf{Reneging \%:} The percentage of EV users who leave the queue after waiting.

\textbf{Service \%:} The proportion of users that are successfully served by charging to the desired SoC, which is the percentage of EVs successfully charged from the total Queued

\textbf{Assumption of Wait Time(AWT):} This is the instantaneous wait time assumption made by the arriving EVs when they observe the state of the CS.

\textbf{Estimated Wait Time(EWT):} This is the wait-time estimation calculated by the CS and shared with the arriving EV users.

\subsection{Results}

In Table~\ref{tab:results}, we list the variation in the percentage of arriving EV users balking and reneging, along with the percentage of users served (by charging to their desired SoC). These percentages are considered the performance metrics of the CS and henceforth will be referred to as balking, reneging, and service metrics.

\begin{figure}[th]
\centering
\includegraphics[width=1\columnwidth]{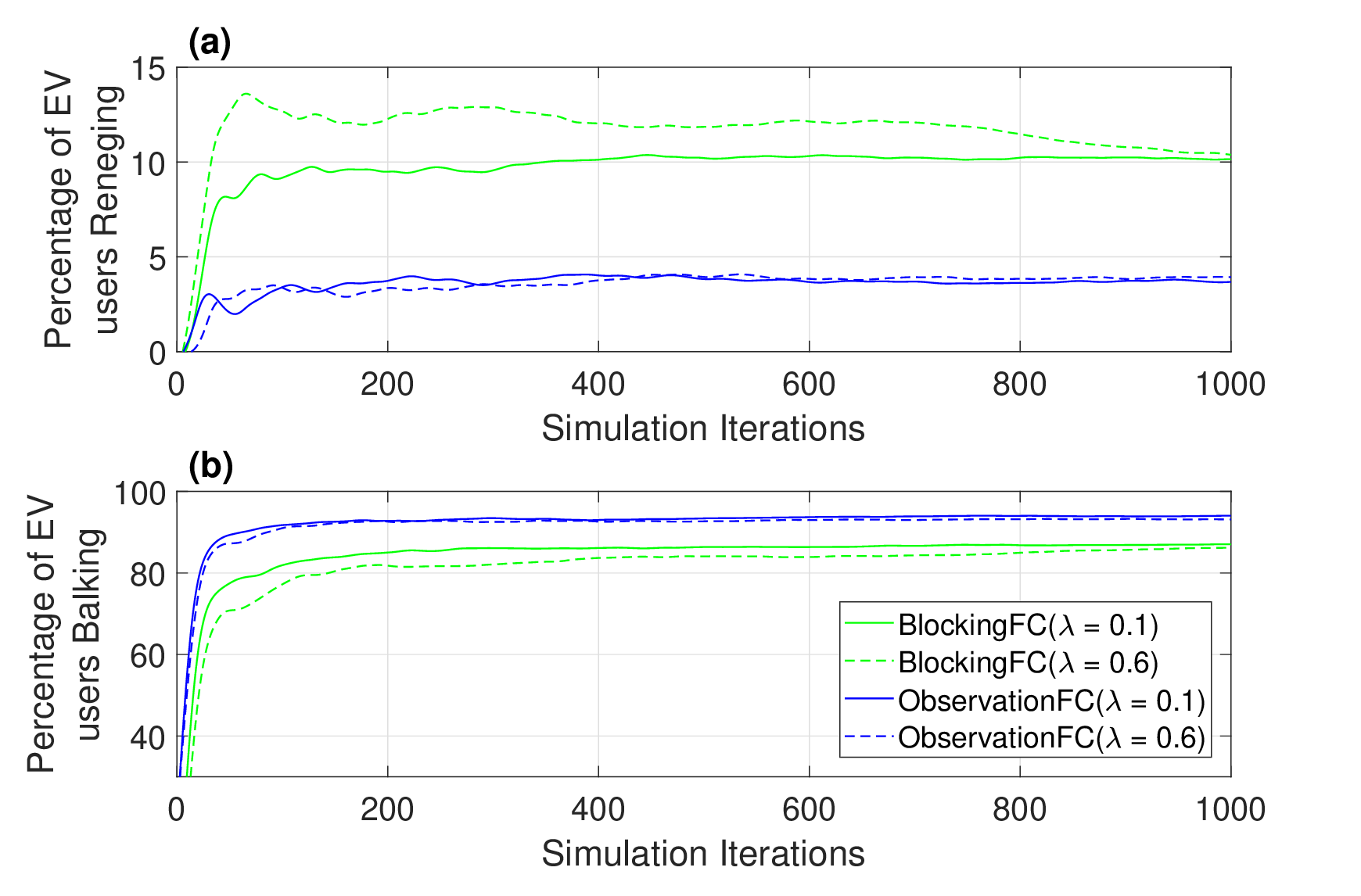}%
\caption{Percentage of EV users (a)Reneging and (b)Balking for cases BlockingFC and ObservationFC for different arrival rates, signifying the charging demand of rush hours and low demand hours in the presence of impatient users reneging}
\label{fig:BalkRenegeObsVsBlock}
\end{figure}

\begin{table}[t]
\caption{Variation in the Percentage of Users Balking and Reneging, with and without station status information sharing\label{tab:results}}
\centering
    \begin{tabular}{|c|c|c|c|c|c|} \hline 
               & Station & Fast      & Balking & Reneging & Service\\
    $\lambda$  & status  & Charger   &   \%    &   \%     &   \%   \\
               & info    & (FC) Type &         &          &        \\ \hline 
        0.1    &  No     & Standard  & 94.01   & 3.67     & 33.06  \\ \hline 
        0.1    &  Yes    & Standard  & 97.33   & 0.31     & 77.19  \\ \hline 
        0.1    &  Yes    & Proposed  & 97.14   & 0.31     & 82.12  \\ \hline
        0.6    &  No     & Standard  & 93.14   & 3.94     & 37.05  \\ \hline
        0.6    &  Yes    & Standard  & 97.09   & 0.35     & 74.57  \\ \hline
        0.6    &  Yes    & Proposed  & 96.72   & 0.35     & 81.01  \\ \hline
    \end{tabular}
\end{table}
The simulation is done anticipating a higher percentage of EV population on the road. We have considered a high arrival rate of $\lambda = 0.6$ and $0.1$, signifying rush hours and normal traffic hours.
Thus, traffic would always be on the higher side in the vicinity of the CS since charging time is significantly higher than the inter-arrival times of the EVs during peak demand hours.
Due to this condition, we would see the balking to be dominated by the forced balking. 
The forced balking cannot be avoided here, as explained before.
In Figure~\ref{fig:BalkRenegeObsVsBlock} we compare the variation of percentages of users reneging and balking for cases {\tt BlockingFC} and {\tt ObservationFC}.
In the {\tt BlockingFC} case, the decision of the EV users does not play a role when joining the queue; hence, the balking is entirely forced balking since the queue is full.
Here, the probability of balking cannot be analytically calculated since we consider an impatient exit here.
Thus, the queue length depends on the service of the charger and the impatience-based exit.
Without the user choice of balking, as in {\tt BlockingFC} a lot of impatient users get admitted to the queue.

It is observed that the Balking is less at rush hours and more otherwise.
This is a very non-intuitive result, which can be inferred from Figure~\ref{fig:BalkRenegeObsVsBlock}.
Since the percentage of EV users reneging is higher during rush hours, the queue is shortened due to higher rate of reneging rather than by faster charging.

\begin{figure}[t]
\centering
\includegraphics[width=1\columnwidth]{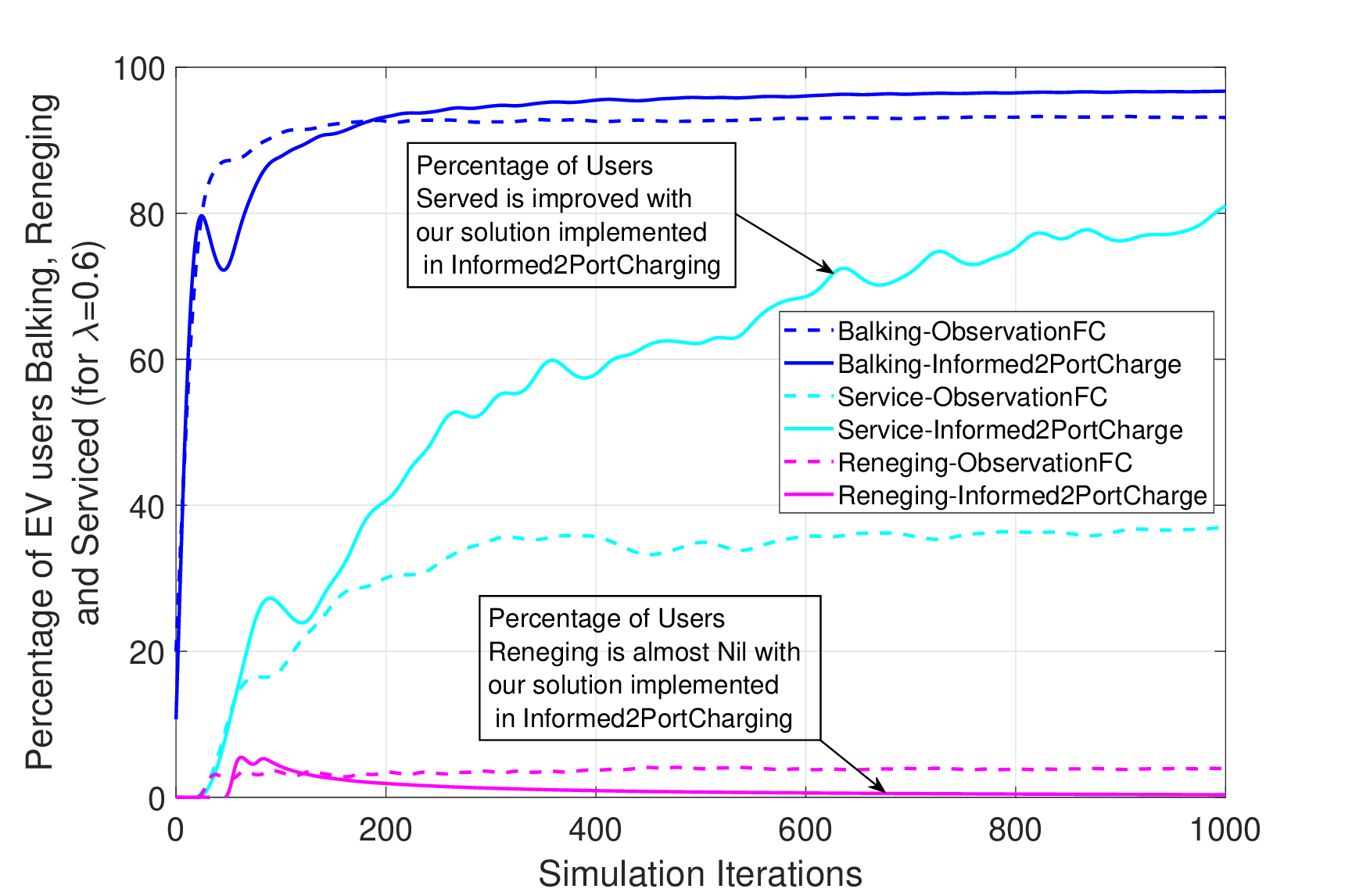}%
\caption{Variation of performance metrics for ObservationFC and Informed2PortCharge. We observe a reduction in users reneging and an increase in the service percentage with our solutions implemented in the case of \textit{Informed2PortCharge}}
\label{fig:AllmetricObsVs2port}
\end{figure}

We see that when the user choice exists in the {\tt ObservationFC} case, the reneging is reduced by 64\% but the balking is increased by 8\%.
The balking now has an additional component of users who choose to balk(voluntary balking).
The scenarios with a lower percentage of user balking do not necessarily ensure a higher percentage of traffic receiving service, as much of the traffic reneges on the queue.
The corresponding improvement in impatience-driven Reneging reduction can be seen in Figure~\ref{fig:AllmetricObsVs2port} with our solution implemented in the case {\tt Informed2PortCharge}.
Here we have shown the variation of the metrics for two different arrival rates for the cases {\tt Informed2PortCharge} compared to {\tt ObservationFC} in Figure~\ref{fig:AllmetricObsVs2port}.
The balking percentage remains higher, mostly dominated by forced balking and the rest from voluntary balking by the user.
The service percentage is the proportion of users that are successfully served by charging to the desired SoC.
The improvement in the case of {\tt Informed2PortCharge} with our proposed solutions implemented can be seen in Figure~\ref{fig:AllmetricObsVs2port}.
This improvement is reflected in a 91\% reduction in the reneging metrics and an 18.72\% increase in the service metrics.
We see that our solution reduces the percentage of users reneging to nil.

\begin{figure}[t]
\centering
\includegraphics[width=1\columnwidth]{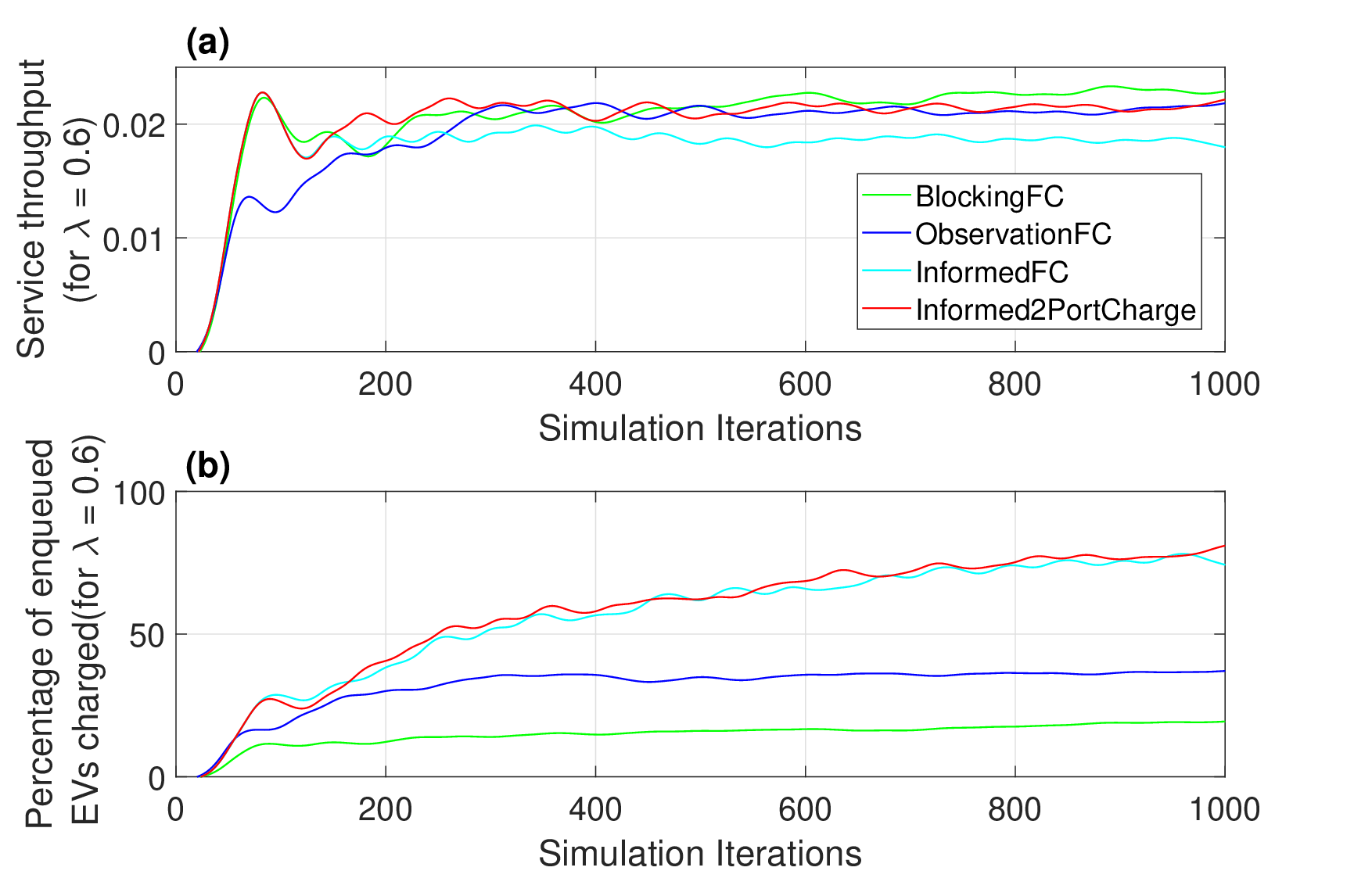}%
\caption{(a)Percentage of EVs Successfully Served from total Queued and (b)Number of EVs Charged per unit Time with EV arrival Rate $\lambda = 0.6$}
\label{fig:ServiceThroughput}
\end{figure}

A relevant metric is the service percentage, which is the percentage of EVs successfully charged from the total queued as shown in Figure~\ref{fig:ServiceThroughput} (a). We compare the service throughput which is the absolute number of EVs served by charging to the desired SoC per unit time in Figure~\ref{fig:ServiceThroughput}(b).
This may not be a quantifiable measure of improvement when considering user impatience.
Throughput is higher when the EVs have a shorter charging demand.
A user with a shorter charging demand has the least patience and leaves the queue, or rather, doesn't join a long queue.
We are left, then, with a queue of users with a longer charging demand. The correct FoM for such a queue is the Service Percentage and not Throughput.

We see a higher throughput for cases {\tt BlockingFC} and {\tt ObservationFC}.
We note that the throughput improved due to our modified charger design implemented in {\tt Informed2PortCharge} exceeding that of the {\tt InformedFC} case.
With our solution, the service percentage increases, keeping the throughput comparable to that of the {\tt ObservationFC} and {\tt BlockingFC} cases during rush hours.

\begin{figure}[t]
\centering
\includegraphics[width=1\columnwidth]{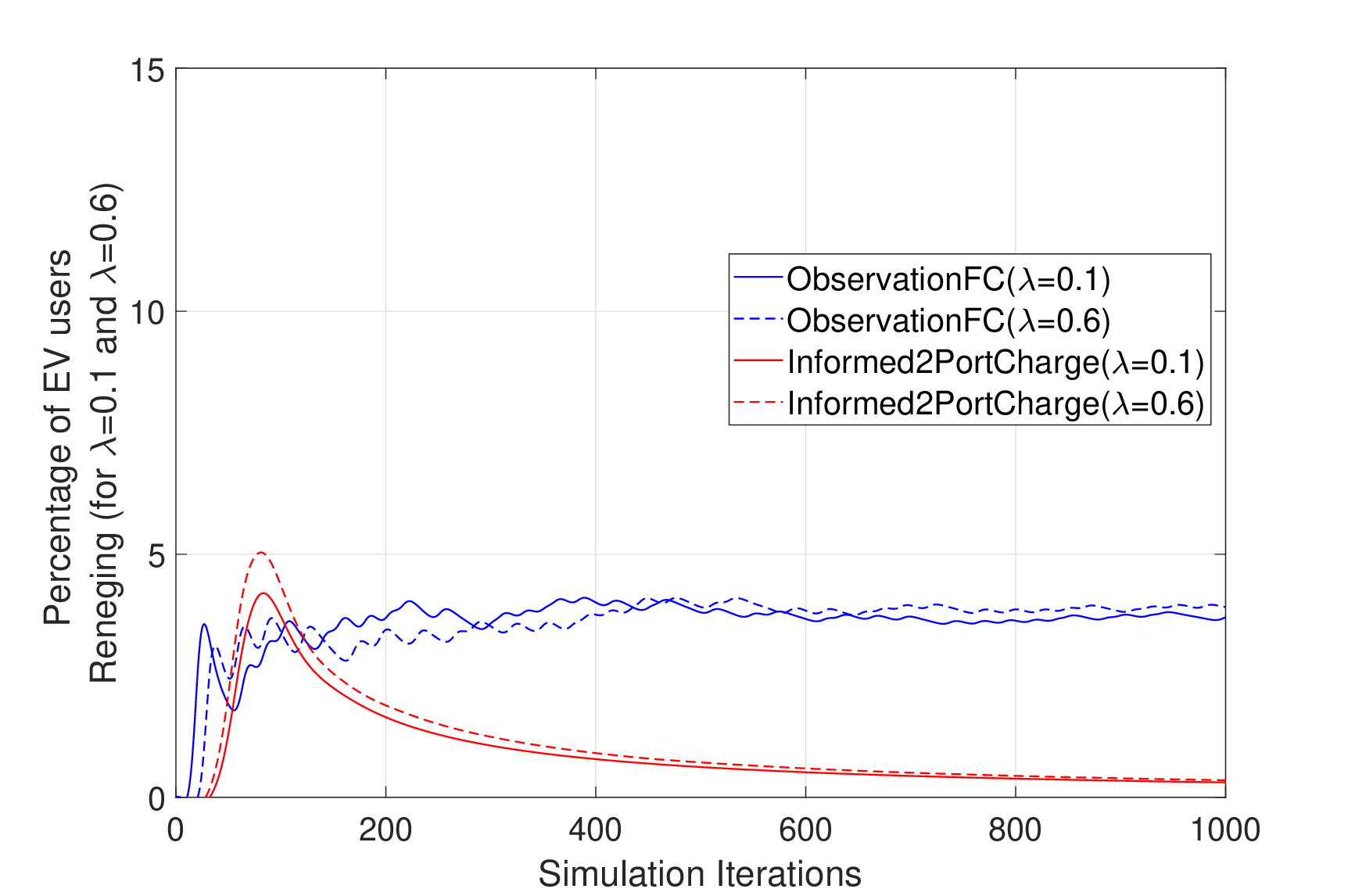}%
\caption{Percentage of EV users Reneging in case of ObservationFC and Informed2PortCharge is observed for traffic arrival rates $\lambda = 0.1,0.6$. Reneging is reduced in the latter case which implements our proposed solution}
\label{fig:RenegeLambda}
\end{figure}

\begin{figure}[t]
\centering
\includegraphics[width=1\columnwidth]{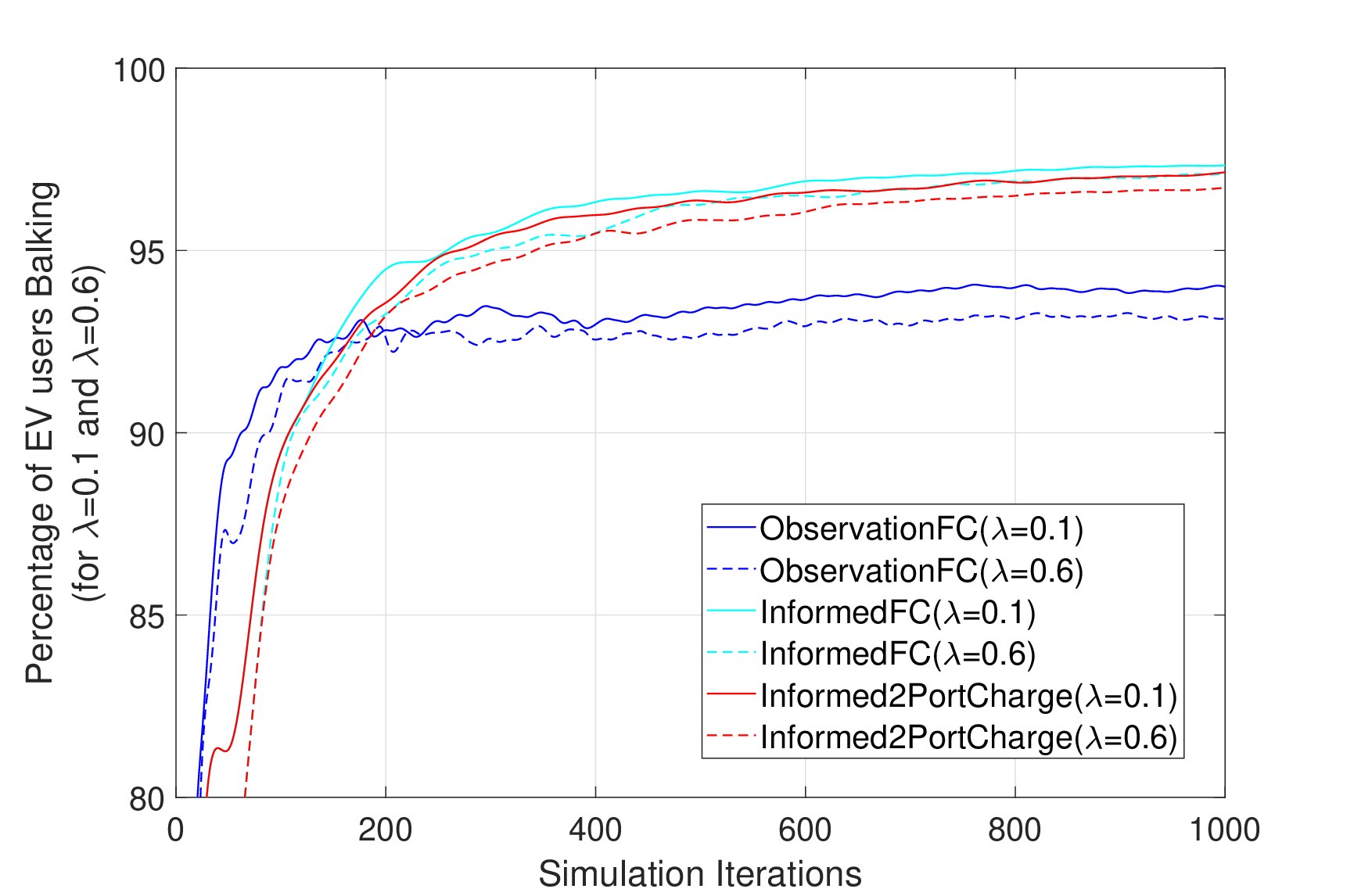}%
\caption{Percentage of EV users Balking in case of ObservationFC, InformedFC and Informed2PortCharge is observed for traffic arrival rates $\lambda = 0.1,0.6$. Balking is relatively lower in the case of Informed2PortCharge compared to InformedFC indicating better charging statistics in the former with 2Port charger.}
\label{fig:BalkLambda}
\end{figure}

We analyze the reneging for different arrival rates(rush hours and normal hours) for the cases {\tt ObservationFC} and Informed2Port charge.
We see in Figure~\ref{fig:RenegeLambda} that the reneging settles near zero for our proposed case while it varies a lot for the case of {\tt ObservationFC}.
When users do not know the Charging Status (from the CS) and forced balking happens, we observe lower traffic loss percentages due to balking but a higher percentage of traffic loss due to reneging users.
Therefore, when uninformed about how much longer the wait is likely to be, though we expect them to play safe by waiting to be charged, a larger number of users renege.
Depending on their specific buffer time (Eqn~\ref{eqn:impatience}), EVs might opt not to join the queue for charging, resulting in Balk decisions.
We compared the percentage of traffic that chose to balk for Cases 1-4 for two different EV arrival rates in Figure~\ref{fig:BalkLambda}.

Balking metrics of {\tt Informed2PortCharge} show a slight improvement over {\tt InformedFC} in equilibrium conditions. 
Interestingly, Balking \textit{reduces} when essential charger status information and average queue waiting times are \textit{not} available.
Un-informed users, therefore, prefer to wait in the queue.
They are averse to the risk of running out of charge.
These users, who assess the wait situation without status information, choose to enter the CS and not balk when the queue length is shorter.
This is more common when the arrival rate is lower ($\lambda=0.1$) as opposed to $\lambda=0.6$ during rush hours.

Interestingly, when status information is available, balking increases.
Users no longer play safe - they risk leaving once they have a better estimate of their waiting time.
Though the traffic lost due to balking is a loss in revenue, it reduces queue congestion when these users renege -- after waiting till they run out of patience.
\begin{figure}[t]
\centering
\includegraphics[width=1\columnwidth]{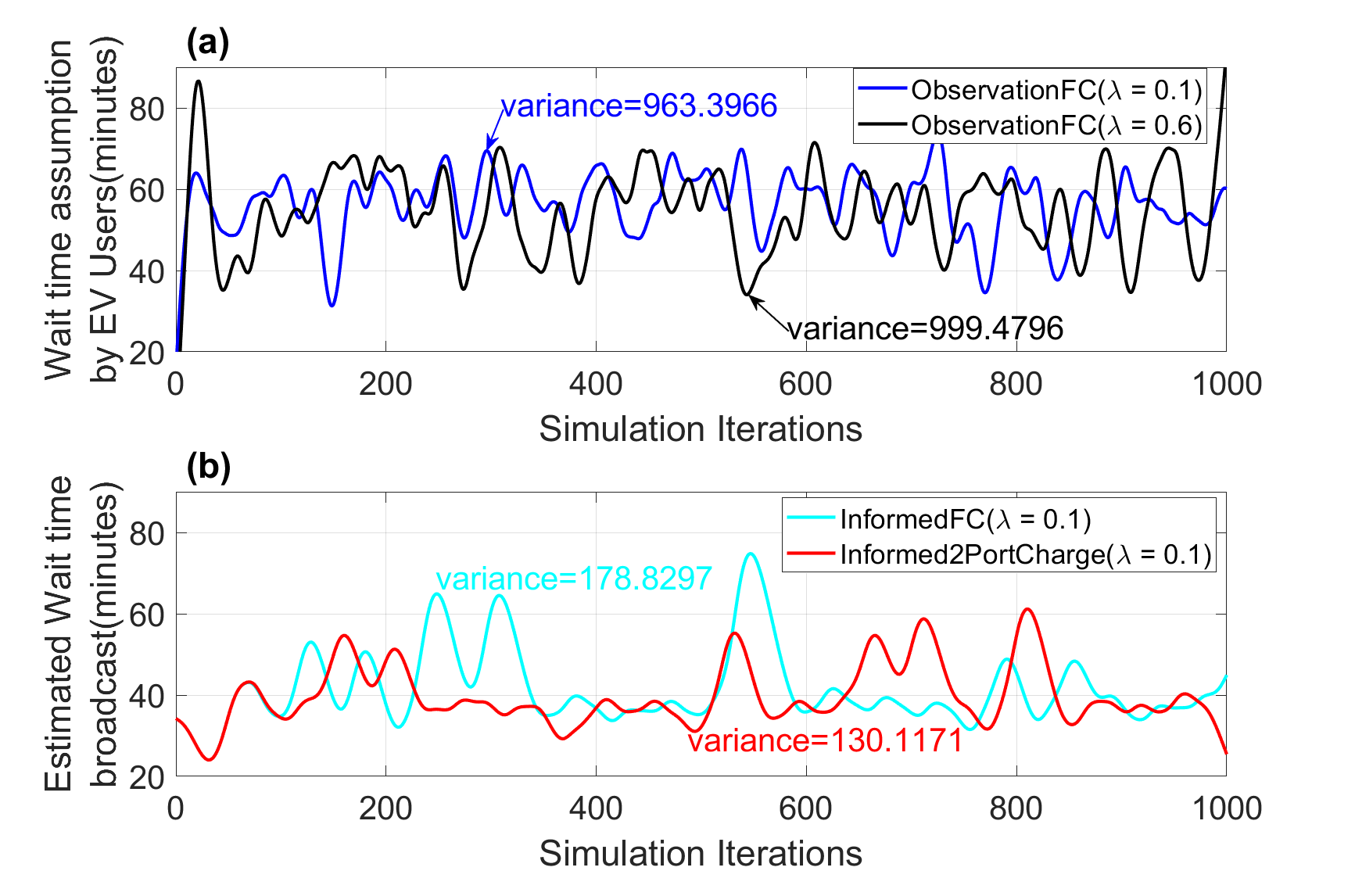}%
\caption{Assumed Wait Time (AWT) by arriving EV users in ObservationFC and the Estimated Wait Time (EWT) data shared with arriving EV users in Informed2PortCharge. The variance is lower in the latter, indicating a stable system and queue}
\label{fig:EstVsAssum}
\end{figure}
The Assumed Wait Time(AWT) and the Estimated Wait Time (EWT) are shown in Figure~\ref{fig:EstVsAssum} where we can see the high variance in AWT as compared to EWT reducing reneging owing to the rational informed decision to join the queue in the first place and stabilizes the queue size.

\begin{figure}[t]
\centering
\includegraphics[width=1\columnwidth]{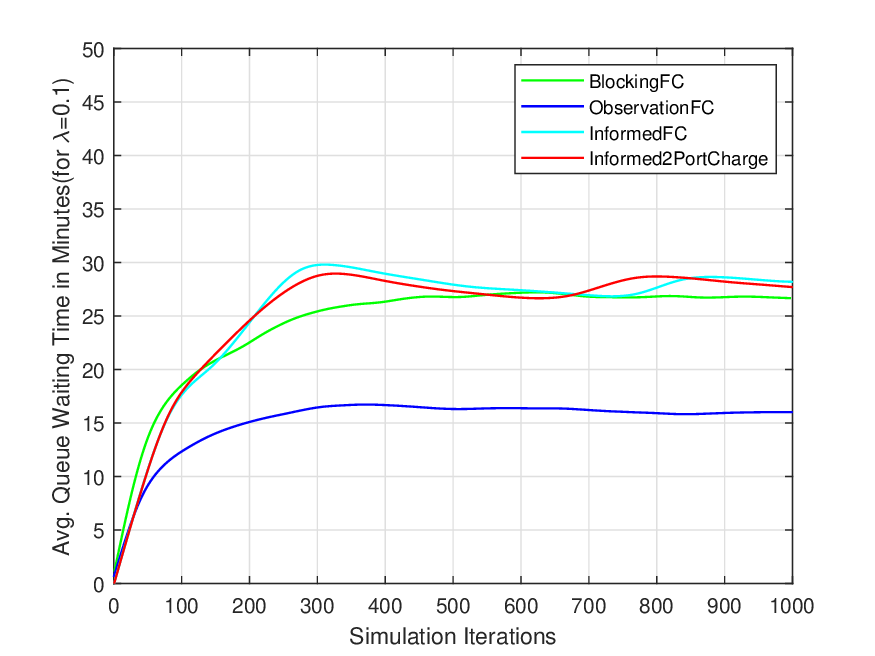}%
\caption{Avg. Wait in the queue for all four cases is observed for EV traffic arrival rate $\lambda = 0.1$. An improvement in Avg. Wait is seen in the case of Informed2PortCharge compared to InformedFC indicating better charging statistics in the former with 2Port charger.}
\label{fig:avg Wait}
\end{figure}

From Figure~\ref{fig:avg Wait}, we observe longer average waiting times for decisions based on {\tt InformedFC}, while cautious users experience shorter waiting times.
This happens because more patient users remain waiting in the queue, their critical battery state requiring longer charging times.
The difference in the waiting time is thus justified.
To enhance the system without aggressive traffic control measures like forced blocking, we devised and tested a modified charger design in {\tt Informed2PortCharge} case.

\section{Conclusion}\label{conclusion}

Our findings highlight the importance of incorporating human factors into CS design and operation for optimal performance.
When considering real-world scenarios with impatient EV user traffic, our evaluation metrics in Section~\ref{Metrics} fare better.
We consider the mostly overlooked lost traffic in Reneging that impacts parking space utilization and CS efficiency.
Our two-pronged solution can increase fast-charger availability by 5\%, improving overall charging efficiency. This is achieved along with a 94\% reduction in demand loss due to reneging EV users ensuring improvement in the quality of service.
This further improves the throughput by 14.28\% and 15.7\% during high demand(rush hours) and low demand respectively.

With extensive simulation through our framework, we have shown how the above design ensures better queue stability and guarantees a QoS for the admitted EVs without compromising on charger utilization, efficiency, or throughput.

Our framework can be further extended by linking it with external tools, as shown in Figure~\ref{fig:layers}. The EVs are modelled as agents having parameters representing both the characteristics of the vehicle along with the behavioural parameters of the user, making it more adaptable to the rapidly evolving EV sector. Besides capacity planning Our framework can be used for policy evaluation and implementation for a mixed type of mobility -- Software Defined Vehicles(SDV), Autonomous EVs, and human-driven EVs incorporating human behaviours.

\section*{Acknowledgments}
We acknowledge the Department of Electrical Engineering, IIT Delhi for generously making available the resources used in accomplishing this work.

\bibliographystyle{IEEEtran}
\bibliography{manuscript_main}

\begin{thebibliography}{10}
\providecommand{\url}[1]{#1}
\csname url@samestyle\endcsname
\providecommand{\newblock}{\relax}
\providecommand{\bibinfo}[2]{#2}
\providecommand{\BIBentrySTDinterwordspacing}{\spaceskip=0pt\relax}
\providecommand{\BIBentryALTinterwordstretchfactor}{4}
\providecommand{\BIBentryALTinterwordspacing}{\spaceskip=\fontdimen2\font plus
\BIBentryALTinterwordstretchfactor\fontdimen3\font minus \fontdimen4\font\relax}
\providecommand{\BIBforeignlanguage}[2]{{%
\expandafter\ifx\csname l@#1\endcsname\relax
\typeout{** WARNING: IEEEtran.bst: No hyphenation pattern has been}%
\typeout{** loaded for the language `#1'. Using the pattern for}%
\typeout{** the default language instead.}%
\else
\language=\csname l@#1\endcsname
\fi
#2}}
\providecommand{\BIBdecl}{\relax}
\BIBdecl

\bibitem{antoun2021data}
J.~Antoun, M.~E. Kabir, R.~F. Atallah, and C.~Assi, ``A data driven performance analysis approach for enhancing the qos of public charging stations,'' \emph{IEEE Transactions on Intelligent Transportation Systems}, vol.~23, no.~8, pp. 11\,116--11\,125, 2021.

\bibitem{brighente2022evscout2}
A.~Brighente, M.~Conti, D.~Donadel, and F.~Turrin, ``Evscout2. 0: Electric vehicle profiling through charging profile,'' \emph{ACM Transactions on Cyber-Physical Systems}, 2022.

\bibitem{TataNexo73:online}
``Tata nexon ev max vs byd e6 : Charging test, curves, graphs - pluginindia electric vehicles,'' \url{https://www.pluginindia.com/blogs/tata-nexon-ev-max-vs-byd-e6-charging-test-curves-graphs}, (Accessed on 12/05/2023).

\bibitem{schwenkMultiDayStochasticScheduling2022}
K.~Schwenk, V.~Hagenmeyer, and R.~Mikut, ``Multi-{{Day Stochastic Scheduling}} of {{Electric Vehicle Charging}} for {{Reliability}} and {{Convenience}},'' in \emph{2022 {{IEEE Vehicle Power}} and {{Propulsion Conference}} ({{VPPC}})}, pp. 1--6.

\bibitem{zhangSimulatingChargingProcesses2022}
Y.~Zhang, R.~Engelhardt, A.-A. Syed, F.~Dandl, C.~Hardt, and K.~Bogenberger, ``Simulating {{Charging Processes}} of {{Mobility-On-Demand Services}} at {{Public Infrastructure}}: {{Can Operators Complement Each Other}}?'' in \emph{2022 {{IEEE}} 25th {{International Conference}} on {{Intelligent Transportation Systems}} ({{ITSC}})}, pp. 2200--2205.

\bibitem{tanQueueingNetworkModels2014c}
X.~Tan, B.~Sun, and D.~H. Tsang, ``Queueing network models for electric vehicle charging station with battery swapping,'' in \emph{2014 IEEE International Conference on Smart Grid Communications (SmartGridComm)}.\hskip 1em plus 0.5em minus 0.4em\relax IEEE, 2014, pp. 1--6.

\bibitem{xiaoOptimizationModelElectric2020a}
D.~Xiao, S.~An, H.~Cai, J.~Wang, and H.~Cai, ``An optimization model for electric vehicle charging infrastructure planning considering queuing behavior with finite queue length,'' \emph{Journal of Energy Storage}, vol.~29, p. 101317, 2020.

\bibitem{esmailiradExtendedQueueingModel2021}
S.~Esmailirad, A.~Ghiasian, and A.~Rabiee, ``An extended m/m/k/k queueing model to analyze the profit of a multiservice electric vehicle charging station,'' \emph{IEEE Transactions on Vehicular Technology}, vol.~70, no.~4, pp. 3007--3016, 2021.

\bibitem{qinChargingSchedulingMinimal2011}
H.~Qin and W.~Zhang, ``Charging scheduling with minimal waiting in a network of electric vehicles and charging stations,'' in \emph{Proceedings of the {{Eighth ACM}} International Workshop on {{Vehicular}} Inter-Networking}.\hskip 1em plus 0.5em minus 0.4em\relax {ACM}, pp. 51--60.

\bibitem{wangElectricalVehicleCharging2018}
S.~Wang, S.~Bi, Y.-J.~A. Zhang, and J.~Huang, ``Electrical vehicle charging station profit maximization: Admission, pricing, and online scheduling,'' \emph{IEEE Transactions on Sustainable Energy}, vol.~9, no.~4, pp. 1722--1731, 2018.

\bibitem{zenginis2016analysis}
I.~Zenginis, J.~S. Vardakas, N.~Zorba, and C.~V. Verikoukis, ``Analysis and quality of service evaluation of a fast charging station for electric vehicles,'' \emph{Energy}, vol. 112, pp. 669--678, 2016.

\bibitem{wang2022queue}
Q.~Wang, D.~Zhang, and B.~Du, ``A queue balancing approach for electric vehicle charging allocation,'' in \emph{2022 IEEE 25th International Conference on Intelligent Transportation Systems (ITSC)}.\hskip 1em plus 0.5em minus 0.4em\relax IEEE, 2022, pp. 2750--2755.

\bibitem{schoenbergReducingWaitingTimes2023}
S.~Schoenberg and F.~Dressler, ``Reducing waiting times at charging stations with adaptive electric vehicle route planning,'' \emph{IEEE Transactions on Intelligent Vehicles}, vol.~8, no.~1, pp. 95--107, 2022.

\bibitem{zhangDiscreteeventHybridTraffic2019}
Y.~Zhang, C.~G. Cassandras, W.~Li, and P.~J. Mosterman, ``A discrete-event and hybrid traffic simulation model based on simevents for intelligent transportation system analysis in mcity,'' \emph{Discrete Event Dynamic Systems}, vol.~29, no.~3, pp. 265--295, 2019.

\bibitem{moghaddamSmartChargingStrategy2018}
Z.~Moghaddam, I.~Ahmad, D.~Habibi, and Q.~V. Phung, ``Smart charging strategy for electric vehicle charging stations,'' \emph{IEEE Transactions on transportation electrification}, vol.~4, no.~1, pp. 76--88, 2017.

\bibitem{zhangDeepLearningBasedProbabilisticForecasting2021b}
X.~Zhang, K.~W. Chan, H.~Li, H.~Wang, J.~Qiu, and G.~Wang, ``Deep-learning-based probabilistic forecasting of electric vehicle charging load with a novel queuing model,'' \emph{IEEE transactions on cybernetics}, vol.~51, no.~6, pp. 3157--3170, 2020.

\bibitem{laiPricingElectricVehicle2023}
S.~Lai, J.~Qiu, Y.~Tao, and J.~Zhao, ``Pricing for electric vehicle charging stations based on the responsiveness of demand,'' \emph{IEEE Transactions on Smart Grid}, vol.~14, no.~1, pp. 530--544, 2022.

\bibitem{ucerModelingAnalysisFast2019}
E.~Ucer, I.~Koyuncu, M.~C. Kisacikoglu, M.~Yavuz, A.~Meintz, and C.~Rames, ``Modeling and analysis of a fast charging station and evaluation of service quality for electric vehicles,'' \emph{IEEE Transactions on Transportation Electrification}, vol.~5, no.~1, pp. 215--225, 2019.

\bibitem{zhangOptimalChargingScheduling2019}
Y.~Zhang, P.~You, and L.~Cai, ``Optimal charging scheduling by pricing for ev charging station with dual charging modes,'' \emph{IEEE Transactions on Intelligent Transportation Systems}, vol.~20, no.~9, pp. 3386--3396, 2018.

\bibitem{bocquet2005queueing}
S.~Bocquet, \emph{Queueing theory with reneging}.\hskip 1em plus 0.5em minus 0.4em\relax DSTO, 2005.

\bibitem{lee_acndata_2019}
Z.~J. Lee, T.~Li, and S.~H. Low, ``{ACN}-{Data}: {Analysis} and {Applications} of an {Open} {EV} {Charging} {Dataset},'' in \emph{Proceedings of the Tenth International Conference on Future Energy Systems}, ser. e-Energy '19, jun 2019.

\bibitem{ucer2019modeling}
E.~Ucer, I.~Koyuncu, M.~C. Kisacikoglu, M.~Yavuz, A.~Meintz, and C.~Rames, ``Modeling and analysis of a fast charging station and evaluation of service quality for electric vehicles,'' \emph{IEEE Transactions on Transportation Electrification}, vol.~5, no.~1, pp. 215--225, 2019.

\bibitem{kleinrock1967time}
L.~Kleinrock, ``Time-shared systems: A theoretical treatment,'' \emph{Journal of the ACM (JACM)}, vol.~14, no.~2, pp. 242--261, 1967.

\bibitem{sztrik2012basic}
J.~Sztrik \emph{et~al.}, ``Basic queueing theory,'' \emph{University of Debrecen, Faculty of Informatics}, vol. 193, pp. 60--67, 2012.

\bibitem{SimEventsUserGuide}
``{{MathWorks. (2021b) SimEvents User}}'s {{Guide (2021b)}}.''

\bibitem{liExtensibleDiscreteEventSimulation2016a}
W.~Li, R.~Mani, and P.~J. Mosterman, ``Extensible discrete-event simulation framework in simevents,'' in \emph{2016 Winter Simulation Conference (WSC)}.\hskip 1em plus 0.5em minus 0.4em\relax IEEE, 2016, pp. 943--954.

\end{thebibliography}
\begin{IEEEbiographynophoto}{Animesh Chattopadhyay}
(Member, IEEE) He received his B.Tech Degree in Electronics and Communication from WBUT, India. Pursued M.Tech Degree from IIT(ISM) Dhanbad. Currently pursuing PhD Degree from IIT Delhi, India. His current research interests span areas of intelligent charging, agent-based modelling, resource optimization, and reinforcement learning.
\end{IEEEbiographynophoto}

\begin{IEEEbiographynophoto}{Subrat Kar}
(Senior Member, IEEE) He is a Professor at the Department of Electrical Engineering since 1994. He holds the Ram and Sita Sabnani Chair Professorship at IIT Delhi and his research areas are in optical communication, switching, access technologies, telecom protocols, embedded systems and high speed networks. He has been with the International Center for Theoretical Physics, Trieste as a Post-Doctoral Fellow (1991-1994). He holds a Doctoral Degree in Electrical Communication Engineering from the Indian Institute of Science, Bangalore (1991).  
\end{IEEEbiographynophoto}
\end{document}